# Hot isostatic pressing of powder in tube MgB$_2$ wires

A. Serquis[1], L. Civale, D. L. Hammon, X. Z. Liao, J. Y. Coulter, Y. T. Zhu, M. Jaime, D. E. Peterson, and F. M. Mueller

Materials Science and Technology Division, MS K763, Los Alamos National Laboratory, Los Alamos, NM 87545

V. F. Nesterenko and Y. Gu

Department of Mechanical and Aerospace Engineering, University of California, San Diego, La Jolla, CA 92093

The critical current density ($J_c$) of hot isostatic pressed (HIPed) MgB$_2$ wires, measured by d.c. transport and magnetization, is compared with that of similar wires annealed at ambient pressure. The HIPed wires have a higher $J_c$ than the annealed wires, especially at high temperatures and magnetic fields, and higher irreversibility field ($H_{irr}$). The HIPed wires are promising for applications, with $J_c > 10^6$ A/cm$^2$ at 5 K and zero field and $> 10^4$ A/cm$^2$ at 1.5 T and 26.5 K, and $H_{irr} \sim 17$ T at 4 K. The improvement is attributed to a high density of structural defects, which are the likely source of vortex pinning. These defects, observed by transmission electron microscopy, include small angle twisting, tilting, and bending boundaries, resulting in the formation of sub-grains within MgB$_2$ crystallites.

---

[1] Electronic mail: aserquis@lanl.gov

One of the best procedures for processing superconducting wires and tapes of the recently discovered[1] superconductor MgB$_2$, is the power-in-tube (PIT) method,[2] which involves filling a metallic tube with superconducting powder and drawing it into a wire. The wire can be subsequently rolled to form a tape. MgB$_2$ has a low superconducting anisotropy, and therefore orienting the crystallographic texture is not essential for high critical current densities. In addition, polycrystalline MgB$_2$ may be free from weak link behavior at grain boundaries, making it easier to achieve good superconducting properties in wires and tapes using traditional PIT methods.

Considerable progress has been made in developing PIT MgB$_2$. Indeed, very high critical current densities J$_c$ (~10$^6$ A/cm$^2$) have been reported at ~5 K and zero field for wires[3,4,5] or tapes.[6,7,8,9] However, these high values rapidly decrease with increasing temperature or magnetic field. In fact, only a few measurements have been reported in the higher temperature range of 20 - 35 K, with one of the best values reported by Soltanian et al.[10] (1.7 x 10$^4$ A/cm$^2$ at 33 K in self-field). For in-field applications it is necessary to obtain higher irreversibility fields, H$_{irr}$. The highest reported H$_{irr}$[3] for PIT MgB$_2$ (~12 T at 5 K) is still below those reported for films or irradiated bulk samples.[9,11]

To improve the performance of MgB$_2$ wires, it is necessary to find a processing method that can introduce more pinning centers and also overcome the poor connectivity between grains, which is primarily caused by porosity. Hot isostatic pressing (HIPing) has been reported to improve superconducting properties in bulk samples[12,13] In particular, a HIPed MgB$_2$ pellet that was was found[14] to have significantly increased J$_c$ at high fields as compared to its un-HIPed counterpart. The improved J$_c$ was attributed to a better intergrain connectivity when the porosity in the MgB$_2$ pellet was eliminated during HIPing. Details on HIPing and properties of samples can be found in Ref. [15].

In this letter, we report the J$_c$, H$_{irr}$ and relevant aspects of the microstructure of annealed and HIPed MgB$_2$ wires prepared by PIT. The HIPed wires exhibit higher J$_c$'s than those of a wire annealed at ambient pressure, especially at high temperatures and magnetic fields. They achieve the highest reported H$_{irr}$ for MgB$_2$ PIT. We correlate the improved performance with the presence of defects, which have adequate size and densities to be relevant pinning centers at high magnetic fields, observed by transmission electron microscopy (TEM).

MgB$_2$ powder was packed into stainless steel tubes (inner and outer diameters were 4.6 and 6.4 mm) and the tubes were cold-drawn into round wires with external diameter of 0.8-1.4 mm and MgB$_2$ core diameter of 0.5-0.9 mm, respectively. For the preparation of HIPed wires, as-drawn wires were cut into 10 cm long pieces and sealed at both ends by electric arc welding. The wires were HIPed at 900°C under an isostatic pressure of 200 MPa for 30 minutes, the pressure was released, and the wires cooled at 5°C/min to room temperature. A HIP cycle, in which the wires were cooled to room temperature before the pressure was released, resulted in wires with good J$_c$'s measured by magnetization but poor transport properties due to the presence of numerous macro-cracks. The depressurization of composite wire may induce macroscopic tensile stresses in MgB$_2$ resulting in macrocracks due to a brittle nature of this superconductor and that elastic properties of steel and magnesium diboride are different.[15] Also, a thermal mismatch between stainles steel and magnesium diboride in general cannot be excluded.

Another wire was heated in vacuum at a fast rate of 35°C/min, maintained at 900°C for 30 minutes, and gas quenched with Ar. In previous work,[4,5] we found that this was an appropriate annealing that can significantly increase $J_c$ of PIT $MgB_2$. This annealing eliminates the micro-cracks produced by the mechanical processing due to a recrystallization process promoted by excess Mg used in the initial powder.

We measured the d.c. transport critical current $I_c$ (using a 1 µV criterion) at 4 and 26.5 K, with the wires immersed in liquid He and Ne, respectively, with the field H applied perpendicular to the wire axis. We have also measured the d.c. magnetization, M, of ~ 4-5 mm long pieces of the same samples, with H parallel to the wire axis, using a Quantum Design SQUID magnetometer. We calculated $J_c$ from magnetization loops, M(H), using the Bean model.

Fig. 1 shows $J_c$ obtained from magnetization, as a function of applied magnetic field in the 5-30 K temperature range for one HIPed wire and for a piece of an appropriately annealed wire (identified as A-2 in Refs. 4 and 5). Also shown in this figure are the transport values measured in liquid Ne at 26.5 K (diamond symbols, solid lines). For both samples, the data at 5 K was corrected for self-field effects, because this effect is important at fields below 1 T. The $J_c(H)$ at 5 K is similar for both samples and the extrapolated zero field $J_c$ is higher than $10^6$ $A/cm^2$. For higher temperatures and magnetic fields, $J_c$ values of the HIPed wire are higher than those of the annealed wire, and the steep $J_c$ drop with H is shifted to higher fields. For both samples, excellent agreement was found between the magnetization and transport values measured in liquid Ne at 26.5 K. These values are among the best reported for this range of temperature in $MgB_2$ PIT.[9]

Fig. 2 displays $J_c$ vs. applied field for the annealed and HIPed wires as measured by magnetization (at 5 K) up to 7 T, and by transport (at 4K) up to 18 T. Again, a good correlation exists between magnetization and transport data indicating that both samples are quite homogeneous. The magnetization $J_c$ for both samples is almost the same for fields below 4 T, but the difference increases with the applied field. The $J_c$ is one order of magnitude higher for the HIPed wire in fields greater than ~12 T. For transport measurements up to 7 T we used samples ~10 cm long in order to eliminate the initial ohmic behavior that we sometimes observed in the I-V curves of shorter wires. However, in the 8-18 T region the wire lengths were constrained to ~3-4 cm long, and these measurements frequently exhibited an initial ohmic behavior. This is a consequence of the poor connectivity between the $MgB_2$ and the stainless steel, which results in a transfer length 1-2 cm long for the current, so some current is always flowing through the stainless steel. This problem progressively disappears as $I_c$ decreases with increasing H. In particular, transfer length and heating problems precluded the collection of reliable data for the HIPed wire below 9 T. In the inset of Fig. 2, the $J_c$ values of our PIT wires are compared with some of the best transport values reported in the literature (Refs. 3, 6, 7 and 8). The $J_c$ of our HIPed wire is one of the best and only the SiC doped $MgB_2$ tape[7] has comparable values, indicating that the defects introduced by HIPing are as effective as SiC doping to enhance the pinning. Besides, the method based on HIPing can easily incorporate doping by SiC or other elements or compounds to provide synergistic effects for enhancement of pinning.

TEM investigations of both wires show[16] that (i) the average density of the HIPed wire is a little higher than the un-HIPed wire, although some porosity is still present, (ii) the MgO at grain boundaries found in the non-HIPed sample, which reduces connectivity among grains, was not present in the HIPed wire, similarly to what we observed in bulk samples,[14] and (iii) a higher density of structural defects exists in the HIPed wire. These three differences contribute to the enhancement of $J_c$ in the HIPed wire as compared to the un-HIPed one. However, the increase in $J_c$ due to the reduction of porosity should be small and field independent. In particular, the high density of defects is the most likely reason for the improved pinning at high fields and for the higher $H_{irr}$. These defects include small angle twisting, tilting, and bending boundaries, which were introduced by the HIPing process, resulting in the formation of sub-grains within $MgB_2$ crystallites.

The $MgB_2$ crystallites normally are of a plate shape with the width on (001) up to 1 μm and the thickness along [001] up to 0.2 μm. Along the [001], the crystallites are usually divided into a few sub-grains with the sub-grain boundaries roughly parallel to (001) plane. The sub-grains can be easily identified in TEM when (001) is parallel to the electron beam direction, as shown in the grains marked with A, B, and C, respectively, in Fig. 3(a). There is usually small angle twisting along [001] and/or tilting around (001) between neighboring sub-grains, resulting in a high density of dislocations distributed in the sub-grain boundaries. Figure 3(b) shows an example of dislocations in a twisting/tilting sub-grain boundary in which black-white contrast appears roughly periodically at an interval of about 2.5 nm, as marked with black arrows, along the boundary. At the center of each black-white contrast point is a dislocation. Two black lines are drawn in Fig. 3(b), with each line parallel to the (001) of sub-grains I and II, respectively, indicating a 5° tilt between the two neighboring sub-grains. The high-resolution image in Fig. 3(b), in which one sub-grain is on a <100> zone-axis while the other is off zone-axis (showing only one-dimensional lattice fringes), suggests some twisting between the two neighboring sub-grains. The higher density of defects in the HIPed sample is obviously induced by the high temperature deformation during HIPing. The HIPing is a complex process that induces high strain deformation at the grain scale, thus creating intra-grain defects. At high H, where the pinning enhancement due to the HIPing is more significant, the distance between vortices is small (~15nm at 8T, ~11nm at 16T), so the defects involved must have a large density. The short separation between the observed dislocations thus makes them good candidates. Dislocations are known to act as pinning centers both in conventional superconductors[17] and in epitaxial $Yba_2Cu_3O_7$ thin films,[18,19] where they also form at low angle grain boundaries.

As we have previously reported,[4,5] appropriate annealing conditions can improve $J_c$ by more than one order of magnitude compared to as-drawn wire, and in this work we show that it also enhances the irreversibility field ($H_{irr}$ > 13 T at 4 K, as shown in Fig. 2). The results for the HIPed wire are even more promising for applications,[20] with high field $J_c$ values around one order of magnitude higher than those of the annealed wire and $H_{irr}$ ~ 17 T (at 4 K). This value is the largest reported in PIT $MgB_2$ wires or tapes, indicating that the defects produced by the HIPing process are effective as pinning centers.

**Figure captions**

**Fig. 1-** $J_c$ vs. applied field for the annealed (open symbols) and HIPed (closed symbols) wires as calculated from magnetization loops in the 5-30 K temperature range. Values measured by transport in liquid Neon at T= 26.5 K (diamond symbols, solid lines) are also included.

**Fig. 2-** $J_c$ vs. applied field for the annealed and HIPed wires as measured by transport (at 4K) and by magnetization (at 5 K). In the inset the Jc values of the same samples are compared with several values reported in the literature (Ref. 3,6,7,8).

**Fig. 3-** (a) Bright-field TEM image of the HIPed wire. Three crystallites, marked with A, B, and C, respectively, clearly show sub-grain boundaries roughly parallel to (001); (b) High-resolution TEM image of a twisting/tilting sub-grain boundary. Black arrows indicate the centers of black-white contrast points at which dislocations are located. Two black lines, each parallel to the (001) of sub-grains I and II, respectively, are drawn together, showing a 5º small angle tilt between the two neighboring sub-grains.

Fig. 1 (A. Serquis et al)

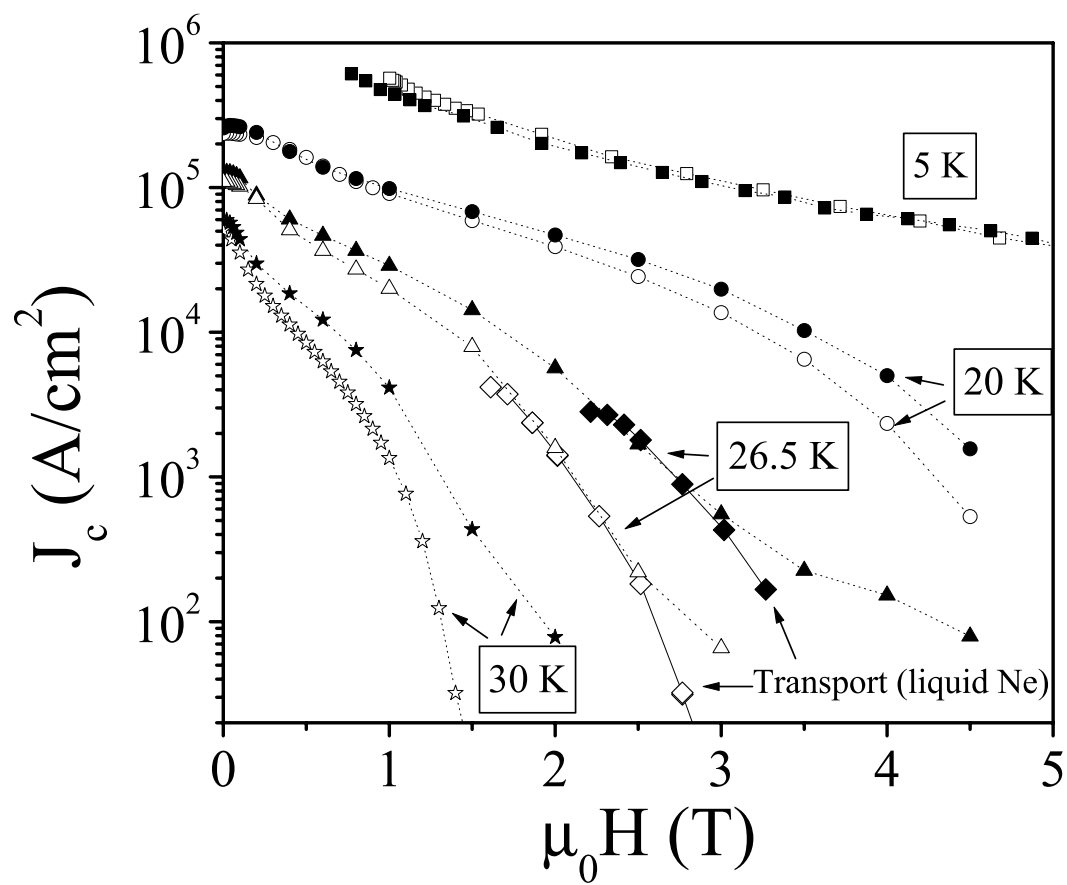

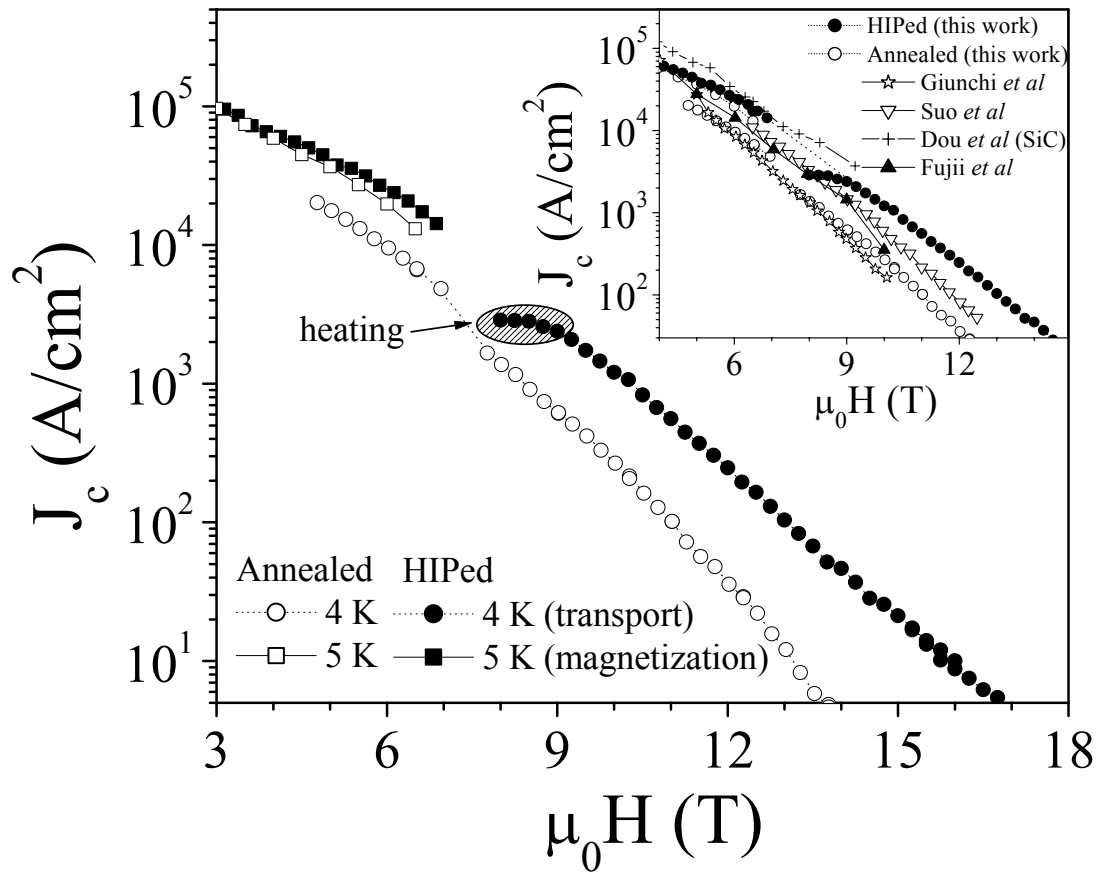

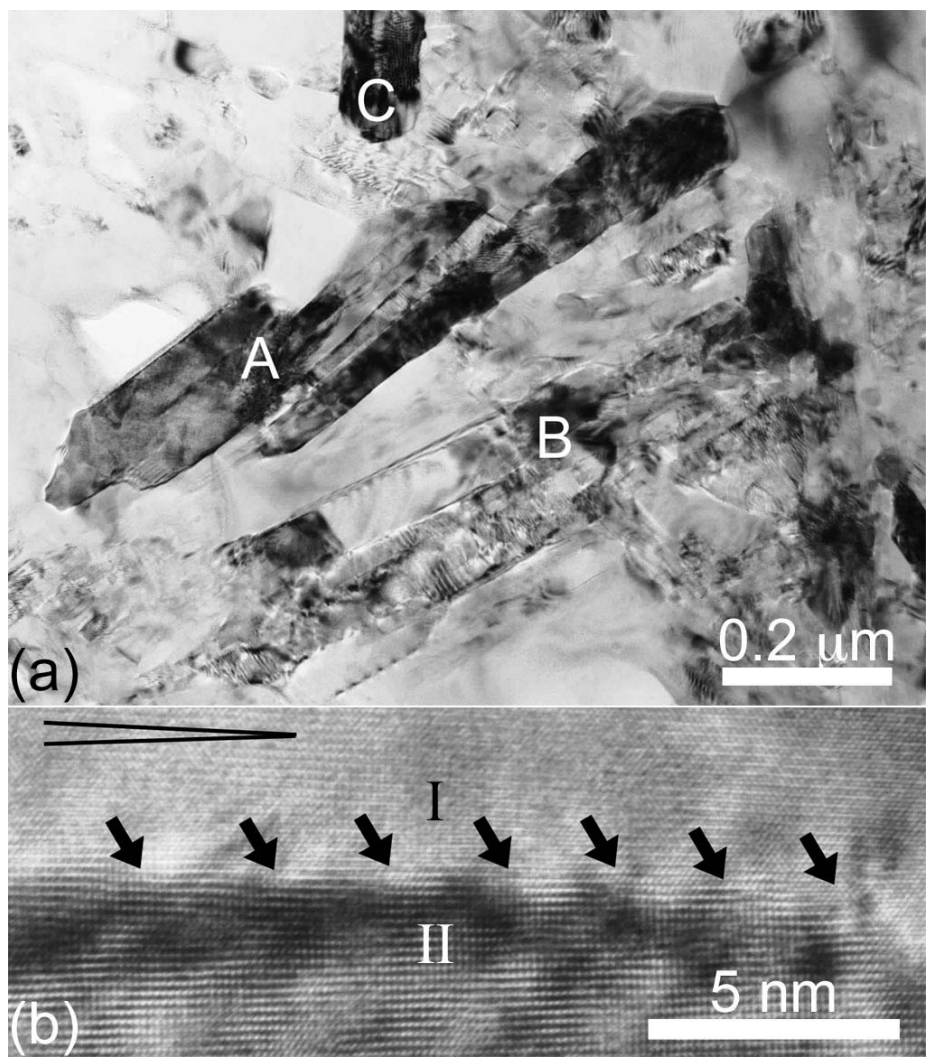

---

[20] HIPing of macroscopically large solenoid type devices with very long wire needed for some practical application is possible with currently available production-sized HIP systems with sizes of workzone about 1 m.